\documentclass[preprint2]{aastex}
\usepackage{graphicx}
\usepackage{amsmath}
\usepackage{amssymb}

\shorttitle{NIR spectroscopy of HD189733b}
\shortauthors{Waldmann et al.}

\begin{document}

\title{Ground-based NIR emission spectroscopy of HD189733b}

\author{I. P. Waldmann, G. Tinetti}
\affil{University College London, Dept. Physics \& Astronomy, Gower Street, WC1E 6BT, UK}
\email{ingo@star.ucl.ac.uk}

\author{P.Drossart}
\affil{LESIA, Observatoire de Paris, CNRS, UniversitŽ Pierre et Marie Curie, UniversitŽ Paris-Diderot. 5 place Jules Janssen, 92195 Meudon France}

\author{M. R. Swain, P. Deroo}
\affil{Jet Propulsion Laboratory, California Institute of Technology, 4800 Oak Grove Drive, Pasadena, California 91109-8099, USA}

\and

\author{C. A. Griffith}
\affil{University of Arizona, Dept. of Planetary Sciences, 1629 E. University Blvd, Tucson, AZ, 85721, USA}

\begin{abstract}


We investigate the K and L band dayside emission of the hot-Jupiter HD~189733b with three nights of secondary eclipse data obtained with the SpeX instrument on the NASA IRTF.  The observations for each of these three nights use equivalent instrument settings and the data from one of the nights has previously reported by Swain et al (2010).  We describe an improved data analysis method that, in conjunction with the multi-night data set, allows increased spectral resolution (R$\sim$175) leading to high-confidence identification of spectral features.  We confirm the previously reported strong emission at $\sim$3.3 $\mu$m and, by assuming a 5$\%$ vibrational temperature excess for methane, we show that non-LTE emission from the methane $\nu_{3}$ branch is a physically plausible source of this emission.  We consider two possible energy sources that could power non-LTE emission and additional modelling is needed to obtain a detailed understanding of the physics of the emission mechanism.  The validity of the data analysis method and the presence of strong 3.3 $\mu$m emission is independently confirmed by simultaneous, long-slit, L band spectroscopy of HD 189733b and a comparison star.

\end{abstract}

\keywords{techniques: spectroscopic, methods: data analysis, planets and satellites: atmospheres, planets and satellites: individual (HD~189733b)}

\section{Introduction}

The field of extrasolar planets is rapidly evolving, both in terms of number of planets discovered and techniques employed in the characterisation of these distant worlds. In recent years, increasing attention has been directed to the detection and interpretation of spectroscopic signatures of exoplanetary atmospheres and was mainly pioneered using Spitzer and HST instruments \citep[eg.][]{agol10, beaulieu08, beaulieu10,charbonneau02,charbonneau05,charbonneau08,grillmair08,harrington06, knutson07b, snellen10b,swain08, swain09a,swain09b,tinetti07,tinetti10}. With the removal of spectroscopic capabilities for Spitzer at the end of the Spitzer cold-phase, increased efforts need to be undertaken to ensure spectroscopic capabilities using ground-based observatories. As inarguably difficult as this is, various groups have succeeded in the detection of metal lines and complex molecules \citep{bean10,redfield08,snellen08,snellen10a,swain10}. 
In order to obtain the desired observations, different groups have developed different techniques. These can be divided into three main categories: 

(1) Time-unresolved techniques: here usually one or more high signal-to-noise (SNR) spectra are taken in and out of transit and both in and out-of-transit spectra are differenced with the additional use of a telluric model. Care needs to be taken to not over-correct and remove the exoplanetary signal.

(2) Time-resolved high-resolution: this is sensitive to very thin and strong emission lines where the exoplanet eclipse is followed with many consecutive exposures and the emission line is identified by the varying doppler shift of the planet as it transits \citep{snellen10a}.

(3) Time-resolved mid-resolution: as above, the exoplanetary eclipse is followed by many consecutive exposures with a mid-resolution spectrograph making this method sensitive to broad roto-vibrational transitions. The use of telluric corrections with a synthetic model is not necessary since we obtain a normalised lightcurve per spectral channel of which the transit depths constitute the spectral signatures. 

Here, we re-analyse the original \citet{swain10} data as well as three additional planetary eclipses  
observed with the IRTF/SpeX instrument. One eclipse, in particular, was obtained with a reference star in the slit. 
We used the time-resolved mid-resolution method pioneered by \citep{swain10} with an improved methodology and data-preprocessing routine. The additional data in conjunction with the more advanced techniques adopted, secured results at higher spectral resolution and smaller error bars. Furthermore, we thoroughly tested our data 
to eliminate/quantify the residual telluric contamination.


\section{Observations and data reduction}

Secondary eclipse data of the hot-Jupiter HD189733b were obtained on the nights of August 11th 2007 (previously been analysed  by \citet{swain10}), June 22nd 2009 and the 12th of July 2009 using the SpeX instrument \citep{rayner03} on the NASA Infrared Telescope Facility (IRTF). The observations were timed to start approximately one to two hours before the secondary eclipse event, until one to two hours post-egress. The instrumental setup was not changed for these three nights. The raw detector frames were reduced using the standard SpeX data reduction package, SpexTool, available for IDL \citep{cushing04}, resulting in sets of 439, 489 and 557 individual stellar spectra for each secondary eclipse event respectively. The extraction was done using the aperture photometry setting with a two arc-second aperture.

In addition we have analysed a fourth secondary eclipse of HD189733b observed on July 3rd 2010 using the same instrument.  As opposed to the other three nights, we observed HD189733b in the L-band only, with a single order, long-slit setting. The one arc minute slit allowed us to simultaneously observe our target and a reference star with a K-band magnitude of 8.05  (2MASS 20003818+2242065).  For not saturating the target star, we kept the exposure time at 8 seconds and employed the standard ABBA nodding sequence throughout the night. Each AB set was differenced to remove the background and the final spectra were extracted using both a custom built routine and standard IRAF routines. We found both extractions to yield the same results but the custom built routine performs better in terms of the final scatter observed. The flux received from the reference star is on average 27 times less than that of the target.

The secondary eclipses in the obtained raw spectra (from here onwards, 'raw' refers to the flat fielded, background corrected, wavelength calibrated and extracted spectra) are dominated with systematic (telluric and instrumental) noise. Consequently, the spectral reduction step is followed by data de-noising and signal amplification steps as described in the following sections.

\section{Extraction of the exoplanetary spectrum}
\label{specextract}

We describe in the following subsections how the planetary signal was extracted from the raw spectra. With the nature of the observations being a combined light (planet and stellar flux) measurement, we employ time-differential spectrophotometry during the time of the secondary eclipse. Standard photometric calibration routines typically achieve a $\sim$1$\%$ level of photometric accuracy, hence further de-noising is necessary to reach the required precision. We first removed the instrument systematics in the data (data cleaning) and then we extracted  the planetary signal in the cleaned data  (spectral analysis).

\subsection{Data-cleaning}
\label{denoise}

To achieve the accuracy we need, a robust cleaning of the data is required. The cleaning process comprises three main steps: 1) Normalising the spectra, getting rid of flux offsets in the timeseries and correcting for airmass variations. 2) Correcting wavelength shifts between spectra by re-aligning all spectra with respect to one reference spectrum. This step removes $\sim80\%$ of outliers. 3) Filtering the timeseries of each spectral channel with adaptive wavelets. This step removes white and pink noise contributions at multiple passbands without damaging the underlying data structure \citet{persival00}. 

\subsubsection{Normalisation}
\label{secnorm}

Firstly, we discarded the spectral information outside the intervals of 2.1 - 2.45$\mu$m and 2.9 - 4.0$\mu$m  to avoid the edges of the K and L photometric bands respectively. Then, we corrected for airmass and instrumental effects. This was achieved in a two step process. We first calculated a theoretical airmass function, $AF = exp (-b \times airmass(t))$,  for each night and divided the data by this function. However, we found this procedure  insufficient since the baseline curvature is caused not only by the airmass but by other instrumental effects (e.g. changing gravity vectors of the instrument). We hence additionally fitted  a second order polynomial to the pre- and post-eclipse baseline of each timeseries and divided each single timeseries by the polynomial. Furthermore,  we normalised each observed spectrum by its mean calculated in a given wavelength band (equation \ref{norm}). 

\begin{equation}
\label{norm}
\overset{\wedge}{F}_{n}(\lambda) = \frac{F_{n}(\lambda)}{\bar{F}_{n}} 
  \begin{cases}
   \lambda = 2.1-2.45 \mu m  & K-band \\
   \lambda = 2.9 - 4.0 \mu m   & L-band
  \end{cases}
\end{equation}

\begin{equation*}
\bar{F}_{n} = \frac{\int_{\lambda_{0}}^{\lambda_{1}}F_{n}(\lambda) \text{d} \lambda}{\lambda_{1} - \lambda_{0}}
\end{equation*}

\noindent where $F(\lambda)$ is the flux expressed as a function wavelength, $\lambda$, for each spectrum obtained, $n$. $\bar{F}_n$ and $\overset{\wedge}F_n(\lambda)$ is the normalised spectrum. In the case of an idealised instrument and constant airmass, the normalisation would be superfluous. However, due to pixel sensitivity variations and  bias off-sets on the CCD chip, the individual spectra need to be normalised to avoid frequent 'jumps' in the individual timeseries. In the domain of the high-interference limit \citep{pag07,swain10}, the astrophysical signal is preserved. We investigated the effects of normalising the spectrum over a whole wavelength band or smaller sub-sections of the spectrum and various combinations of both, but found the differences to be negligible.

\subsubsection{Spectra re-alignment and filtering}

\begin{figure}
\epsscale{1.0}
\plotone{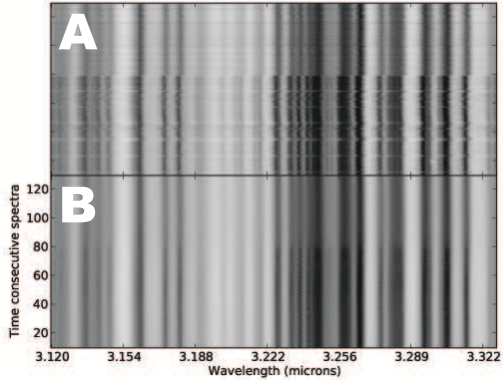}
\caption{Zoomed in fraction of the data prior to the cleaning process (A) and post cleaning (B). Each column is a timeseries at a specific wavelength and each is an individual spectrum ($n$) taken at a specific time. \label{fig1}}
\end{figure}

After the normalisation, we constructed 2D images with rows representing spectra of the planet-star system at a specific time, and columns representing timeseries for specific wavelengths (see figure \ref{fig1}A). 
In figure \ref{fig1}A, the main sources of outliers in individual timeseries, are miss-alignments  by up to 4 pixels along the wavelength axis. We corrected this effect by fitting Gaussians to thin (FWHM $\sim$ 5px) emission and absorption lines to estimate the line centres to the closest pixel. When the shift occurred for all the lines, the spectrum was corrected with respect to a reference spectrum, i.e. the first spectrum in the series. Then cosmic rays were removed by a 2D median filter replacing 5$\sigma$ outliers with the median of its surrounding 8 pixels. 

\subsubsection{Wavelet de-noising}
\label{secwave}

Due to variations in detector efficiency, the cumulative flux of each spectrum depends on the exact position of the spectrum on the detector (horizontal bands in figure \ref{fig1}A), resulting in  high frequency scatter in each individual timeseries. This effect was already attenuated by the normalisation step but further removal of systematic and white noise  is required. Based on the de-noising approach proposed by \citet{thatte10}, we have opted for a  wavelet filtering of the individual timeseries using the 'Wavelet Toolbox' in MATLAB. There are clear advantages to wavelet de-noising compared to simple smoothing algorithms. With wavelets we can specifically filter the data for high frequency 'spikes' and low frequency trends without affecting the  astrophysical signal or losing temporal phase information. This allows for an efficient reduction of white and pink noise in the individual timeseries. By contrast, smoothing algorithms, such as kernel regression, will impact the desired signal since these algorithms smooth over the entire frequency spectrum \citet{donoho95} and \citet{persival00}. For a more detailed discussion see appendix and \citet{thatte10, donoho95, persival00, stein81, sardy00}. The use of the wavelet filtering to each individual timeseries yielded a factor of two improvement on the final error bars. The final results were generated 
with and without wavelet de-noising and found to be consistent within the respective errorbars. An example of the final de-noised data can be seen in figure \ref{fig1}B.

\subsection{Measuring the exoplanetary spectrum}
\label{fourier}

After the data were de-noised as described in the previous subsection, we focused on the extraction of the planetary signal. We based our analysis on the approach described in \citet{swain10}. The spectral emission features of a secondary eclipse event are too small to be statistically significant for an individual spectral channel. High signal to noise detections require a low spectral resolution, i.e. binning the data in $\lambda$. This can be done more efficiently  in the frequency domain for reasons discussed below. Each timeseries $X_i(t)$ (here $i$ denotes the spectral channel) was re-normalised to a zero mean to minimise windowing effects in the frequency domain. The discrete fast-Fourier transform (DFT) was computed for each timeseries and, depending on the final binning,  $m$ number of Fourier-transformed timeseries were multiplied with each other and finally normalised by taking the geometric mean (equation \ref{equ1}). 

\begin{equation}
\label{equ1}
\mathcal{F}[ \bar X(t)] = \left( \prod_{i=1}^{m} \mathcal{F} [ X_{i}(t) ] \right) ^{1/m}
\end{equation}

\noindent where $\mathcal{F}$ is the discrete Fourier-transform and $X_{i}(t)$ is the timeseries for the spectral channel $i$ for $m$ number of spectral channels in the Fourier product ($m \in  \mathbb{Z}^+$). Since the input timeseries are always real and the Fourier transforms are Hermitian, we can take the n'th root of the real-part of the final product without loosing information.

In the time-domain, this operation is equivalent to a consecutive convolution of $X_i$ with $X_{i+1}$, equation \ref{conv2}.

\begin{equation}
(X_{i} \ast X_{i+1})[n] \overset{def}{=} \sum_{t=1}^{n} X_{i}[t]X_{i+1}[n-t]
\label{conv2}
\end{equation}

We can appreciate from equation \ref{conv2} that one eclipsing timeseries is the weighting function of the other. The consecutive repetition of this process for all remaining ($i$-1) timeseries, effectively filters the convolved timeseries with the weighting function that is another timeseries. 
This has the effect of smoothing out noise components whilst preserving the signal common to all the timeseries sets \citep{pag07}. 
The final result of this process is the geometric mean of all timeseries. 
For an individual timeseries, the eclipse signal may not be statistically significant but the simultaneous presence of the eclipse in all the timeseries allows us to amplify the eclipse signal to a statistical significance by suppressing the noise. The {\it convolution theorem} states that the Fourier transform of a convolution is equivalent to the dot product of the Fourier transforms

\begin{equation}
 \mathcal{F}(X_i \ast X_{i+1}) \equiv k \otimes \mathcal{F}(X_i) \otimes \mathcal{F}(X_{i+1}) 
 \label{dotpro}
 \end{equation}
 
 \noindent where $\otimes$ signifies multiplication in the Fourier space and  $k$ is a normalisation constant.  This process is the base of the our analysis.

\subsubsection{Time-domain analysis}
\label{timedomain}

Calculated the Fourier product, $\mathcal{F}[ \bar X(t)]$, for $i$ spectral channels, we can take the inverse of the Fourier transform to obtain the filtered lightcurve signal.  

\begin{equation}
\bar X(t) = \mathcal{F}^{-1}(\mathcal{F}[ \bar X(t)])
\label{totime}
\end{equation}

\noindent The  lightcurves were then re-normalised by fitting a second-order polynomial to the out-of-transit baseline. We modeled the  final lightcurves with equation 8 of \citet{mandel02}, using the system parameters reported in \citet{bakos06}, with the transit depth as the only free parameter left. 

As clear from the lightcurves presented in section \ref{results}, the systematic noise  in the data is  higher in areas of low transmissivity. Systematic noise increases the scatter of the obtained lightcurves as well as the error-bars of the final spectra and places a lower limit of $m$ =  50 channels ($\sim 2.88$nm) on the currently achievable spectral bin size. This is a noticeable improvement compared to the original \citet{swain10} analysis which reported a lower limit of $m$ = 100  and 150 spectral channels for the K and L-bands respectively.

\subsubsection{Frequency-domain analysis}
\label{frequdomain}

The generated lightcurves  are of high quality and ready for accurate spectroscopic measurements. However, as previously mentioned, a certain amount of periodic and systematic noise is still present  in the timeseries. The noise residuals  are in part generated during  the conversion of the data from the frequency domain to the time domain, in part are due to systematics. We can remove some of these residuals, by measuring the  eclipse depth directly in the frequency domain, assuming that most systematic noise is found at different frequencies to the eclipse signal. 

In first order approximation, we can assume the eclipse signal to be a box-shaped function or square wave of which the Fourier transform is the well known sinc function \citep{riley04}. The Fourier series of such a symmetric square wave is given by equation \ref{sync} as a function  of the lightcurve's transit depth, $\delta$, and transit duration, $\tau$.

\begin{align}
\label{sync} 
f(t) &= 4\delta\left(\text{sin}(2 /\tau) - \frac{1}{3} \text{sin}(3 /\tau) + \frac{1}{5} \text{sin}(5/\tau)-  ... \right) \\\nonumber
&= 4\delta \sum_{k=1,3,5...}^{\infty}\frac{\text{sin}((2k-1)2 /\tau)}{(2k-1)}
\end{align}

\begin{figure}
\epsscale{1.0}
\plotone{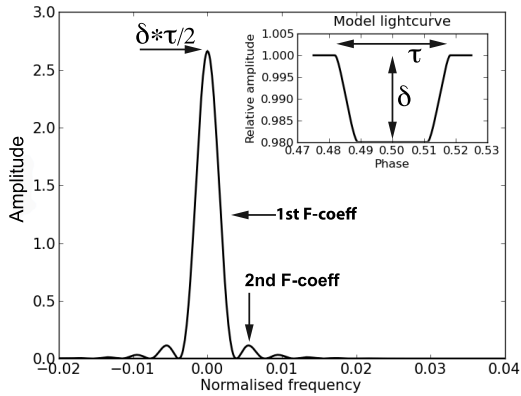}
\caption{showing power spectrum of a \citet{mandel02} model lightcurve of HD189733b (inset). It can clearly be seen that most power of the lightcurve signal is contained in the first Fourier coefficient. \label{fig2}}
\end{figure}

The lightcurve signal is composed of a series of discrete frequencies, $k$, since the boundary conditions of the function are finite. This series is very rapidly converging. Figure \ref{fig2} illustrates this. Here we took the Fourier transform of the secondary eclipse model shown in the insert. The frequency spectrum is centred on the first Fourier coefficient. It is clear that most of the power is contained in the first Fourier coefficient and the series quickly converges asymptotically to zero after the third coefficient. Taking the product in equation \ref{equ1} has the effect of strengthening the eclipse signal, whilst weakening the noise contribution: the frequencies contributing to the noise  are in fact expected to be different to the ones
contributing to the eclipse signal. In the case of stochastic (Gaussian) noise, wavelength dependent instrumental noise or scintillation noise, this is obvious.

Following Fourier series properties, the modulus of the amplitude, $|A|$, of the coefficients in equation \ref{sync} is directly proportional to the transit depth $\delta$ and the transit duration $\tau$, where $\tau = t_{1-4} / t_s$ and $t_{1-4}$ is the transit duration from the first to fourth contact point and $t_{s}$ is the sampling rate (ie. exposure time + overheads, fig. \ref{fig2}). 

 \begin{equation}
|A|_{sqrwave} = \frac{\tau \delta}{2} \sum_{k=1,3,5...}^{\infty} \frac{1}{(2k-1)} 
\label{equ2} 
\end{equation}

The amplitude of the Fourier coefficients above $k=1$ decreases by $(2k-1)$ for a box-shape function and is an even faster converging series for real lightcurve shapes which are used in the analysis (see appendix).
Following from equation \ref{equ2} we see that for the first Fourier coefficient, $k = 1$, the relationship between the transit depth, $\delta$, and the Fourier coeffcient amplitude, $|A|$, is simply given by $|A_{k=1}| = (\tau /2) \delta$. From the analytical arguments presented above, we know that $\tau$ is the transit duration (in units of number of observed spectra). We checked the consistency of the theory with the data, by calculating the value of $\tau$ numerically. To calculate $\tau$ we produced secondary eclipse curves with the transit duration and sampling rate of the original IRTF data sets \citep[equation 8]{mandel02}. We generated 300 curves with transit depths ($\tau$) ranging from 0.0001 to 0.1 and measured the corresponding amplitude ($|A_{k=1}|$). Here, the derivative, $d(|2A|)/d\delta$ gives us the value of $\tau$. We find $\tau$ = 116 in-eclipse measurements, which agrees with the number of in-transit spectra obtained for the real IRTF data-sets. 
 
$N$ spectra were obtained at a constant sampling interval of $t_s$, giving us a sampling rate of $R = 1/t_s$ in the frequency domain. For a complete representation of the data, the sampling rate is equal to the Nyquist rate, $R = 2B$, where B is the spectral bandwidth of the Fourier transform. The total number of Fourier coefficients, $K$, is then given by $K = 2BN$. It follows that the resolution in the frequency domain is determined by $\Delta f = 1/N$. In other words, the more measurements are available the more Fourier coefficients can be extracted to describe the data and consequently the frequency range covered by each coefficient is smaller for a fixed sampling rate. 

The fact that   $\Delta f$ is finite ($\Delta f \rightarrow 0$ for infinitely sampled data-sets), means that the first Fourier coefficient can be contaminated by remaining noise signals very similar in frequency. To estimate the error bar on this contamination, we varied the out of transit (oot) length $N_{oot}$ by 50$\%$ and calculated the resulting spectrum for each $\Delta f$. The error is then estimated as the standard deviation to the mean of all computed spectra. 


\subsection{Application to data}

We have applied the same procedure described in sections \ref{denoise} \& \ref{fourier} to  the four data sets.  In addition to the individual analysis, we also combined in the frequency domain the three  data sets recorded with the same observational technique.  Given that the  low-frequency systematics --such as residual airmass function, telluric water vapour content, seeing, etc-- are significantly different for each individual night, by combining multiple data sets, we  can amplify the lightcurve signal and reduce the systematic noise.

To generate the final K and L-band spectra, we chose  in equation \ref{equ1}  $m=100$ spectral channels. From $R_{spectra} = \lambda_{centre} / \Delta \lambda$, we get  a final spectral resolution of R $\sim$ 50. Combining all three data-sets together ($\sim$33 spectral channels taken from each  observed
planetary eclipse) we obtain a spectral resolution of  $\sim$170 and $\sim$ 185 for the K and L-bands respectively. We note that the  spectral resolving power  for the SpeX instrument, considering the seeing, is R $\sim$ 800.

\section{Model}
\label{non-LTE}

We have simulated planetary emission spectra using lineby-line radiative transfer models as described in \citet{tinetti05,tinetti06} with updated line lists at the hot temperatures from UCL ExoMol and new HITEMP \citep{barber06,yurchenko11,rhotman09}. Unfortunately accurate line lists of methane at high temperatures covering the needed spectral range are not yet available. We combined HITRAN 2008 \citep{rhotman09}, and the high temperature measurements from \citep{thievin08}. These LTE-models were fitted to the spectra presented in section \ref{results}. 

Additional to the standard LTE model, we considered possible non-LTE models to fit the presented data. Upper atmospheres of planetary atmospheres are subject to non-LTE emissions; although negligible in most part of the near infrared spectrum, these emissions become dominant in the strongly absorbing vibration bands of molecular constituents, like CO$_2$ in telluric planets and CH$_4$ in giant planets (and Titan). 
A synthetic model of the spectrum in the L band has been adapted from a model of Giant Planets fluorescence of CH$_4$ developed for ISO/SWS \citep{drossart99}. The main steps involved in the radiative transfer with redistribution of frequency in non-LTE regime can be summarised as  follows:

\begin{itemize}
\item We first calculate the solar (stellar) flux absorbed
from all bands of CH$_{4}$. Although classical, this part of the model can be cumbersome as all the main absorption bands corresponding to the stella flux have to be (in principle) taken into account. Limitations come from the knowledge of the spectroscopy of the hot bands. In this model, the following bands are taken into account: Pentad (3.3 micron) Octad (2.3 micron); Tetradecad (1.8 micron). An estimate of the accuracy of the approximation in neglecting hotter bands will be given below.
Following an approach given by \citet{doyenette98}, the spectroscopy of CH$_{4}$ is simplified by dividing the vibrational levels in stretching and bending modes: therefore x superlevels (instead of the 29 potential sub-levels of the molecule). It is also assumed that for each super-level belonging to a polyad, thermal equilibrium is achieved within the population. This assumption comes from the observation that intra-vibrational transitions within polyads have a higher transition rate than inter-vibrational transitions.
\item The population of the vibrational levels is then calculated within each "super-level" of CH$_4$.
The vibrational de-excitation is assumed to follow the bending mode de-excitation scheme \citep{appleby90}.
\item 
From the population of the each super-level, the
radiative rate of each level can be calculated to determine
the emission within each of the bands (fundamental, octad-dyad and tetradecad-pentad) that contributing to the 3.3 micron domain.
\item If hot band emission can be proven to remain optically thin down to deep levels of the atmosphere, the resonant fluorescence is not the same, as self-absorption is an essential ingredient of the fluorescence. Evidently, photons absorbed, on average, at a tau=1 level have the same probability to be re-absorbed as re-emitted upwards. The optically thick fluorescence, including absorption and re-emission, is therefore applied to the resonant band.
\end{itemize}


\section{Results}
\label{results}
\subsection{Validation of the method used} 
As described in previous sections, we analysed four nights of observations: three in multi-order mode, with only HD~189733b in the slit (referred to as 'short-slit nights') and one night in L-band with  single order, long-slit set up, observing HD~189733b and a fainter reference star simultaneously. While the long-slit observation covers a narrower spectral interval compared to the other eclipse observations, it 
is a critical test of the methodology with its simultaneous observations of the target and the reference star. In figure \ref{mstar2} we present two lightcurves:  HD~189733 and the reference star. Both are centred at 3.31$\mu$m with a binning width of 50 channels ($\sim 2.88$nm). As expected the HD~189733 timeseries (top) shows the distinctive lightcurve shape whilst the reference star (bottom) timeseries shows a null result. We have fitted a \citet{mandel02} secondary eclipse lightcurve to both and found the HD~189733b transit depth to be $\delta_{HD189} = 0.0078 \pm 0.0003$ and $\delta_{REF} = 0.0 \pm 0.0007$ respectively. These results are in good agreement with the  spectra presented below. 
\subsection{K and L-band spectra}
\begin{figure}
\epsscale{1.0}
\plotone{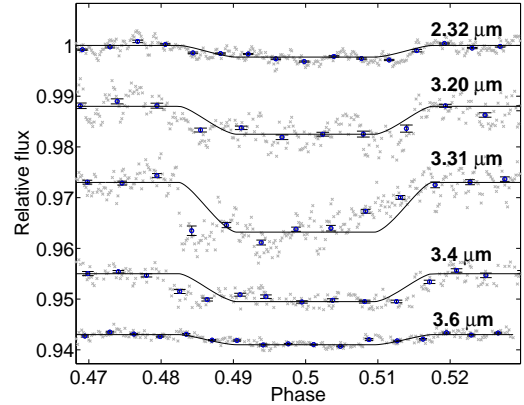}
\caption{Lightcurves of the 'three-night-combined' analysis for the K and L bands. Lightcurves are offset vertically for clarity.  \label{allcurves}}
\end{figure}

\begin{figure}
\epsscale{1.0}
\plotone{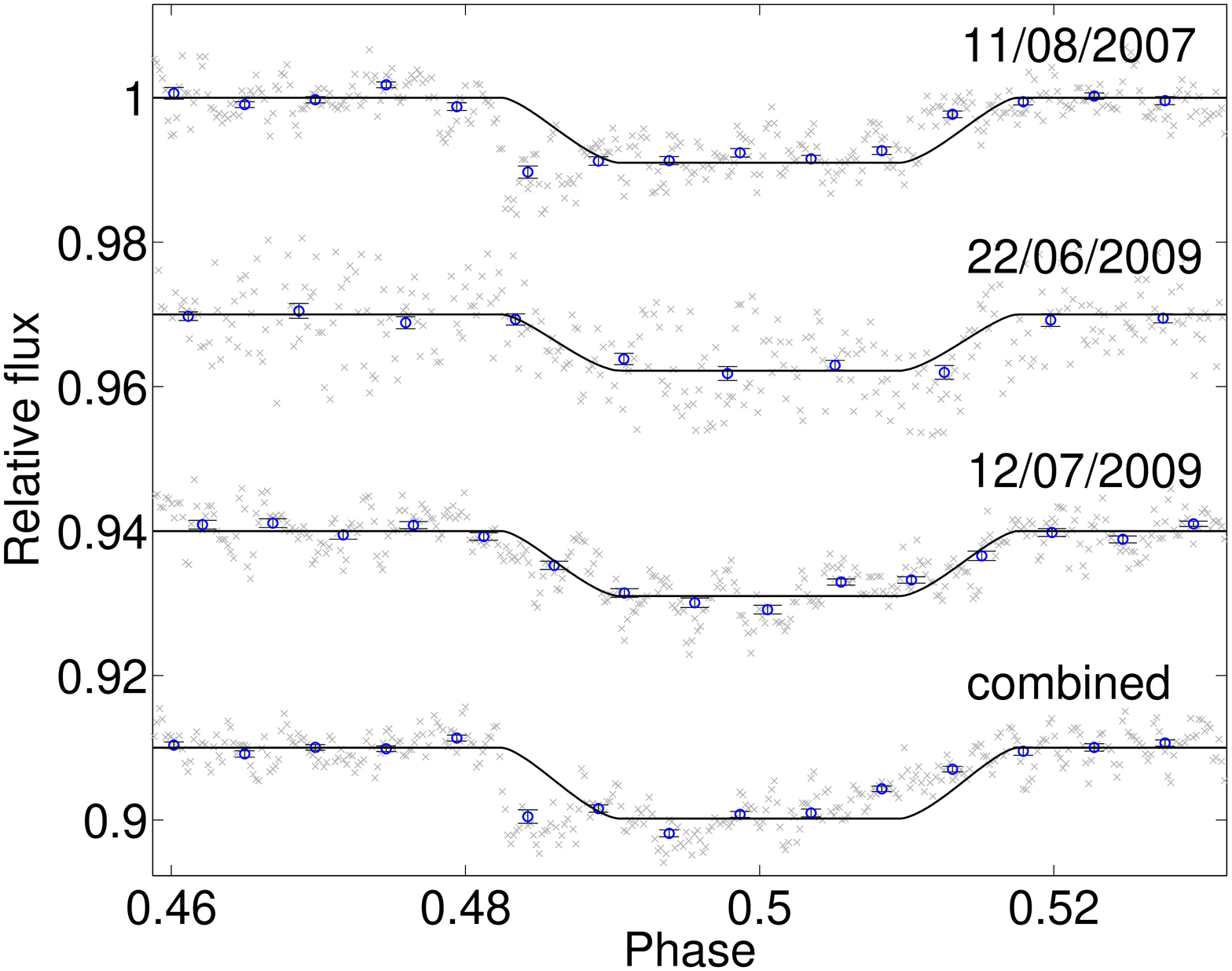}
\caption{Lightcuves centred at 3.31$\mu$m with a bin size of 50 channels ($\sim2.88$ nm) for the three individual nights and 'three-nights-combined'.  \label{3nightnu3}}
\end{figure}

The same analysis was undertaken for the three short-slit nights: illustrative lightcurves are presented in figures \ref{allcurves} \& \ref{3nightnu3}. In figure \ref{allcurves} are plotted the  lightcurves of the 'three-nights-combined' analysis for the K and L-band bands centred at 2.32, 3.20, 3.31 3.4 and 3.6 microns, with 50 channel ($\sim 2.88$ nm) bins. The residual systematic noise   is most pronounced in the areas of low atmospheric transmissivity, which is reflected in the error bars of the lightcurves and of the retrieved spectra. We also
show the lightcurves centred on the methane $\nu_{3}$ branch at $\sim$3.31$\mu$m for all individual nights, figure \ref{3nightnu3}.  

\begin{figure}
\epsscale{1.0}
\plotone{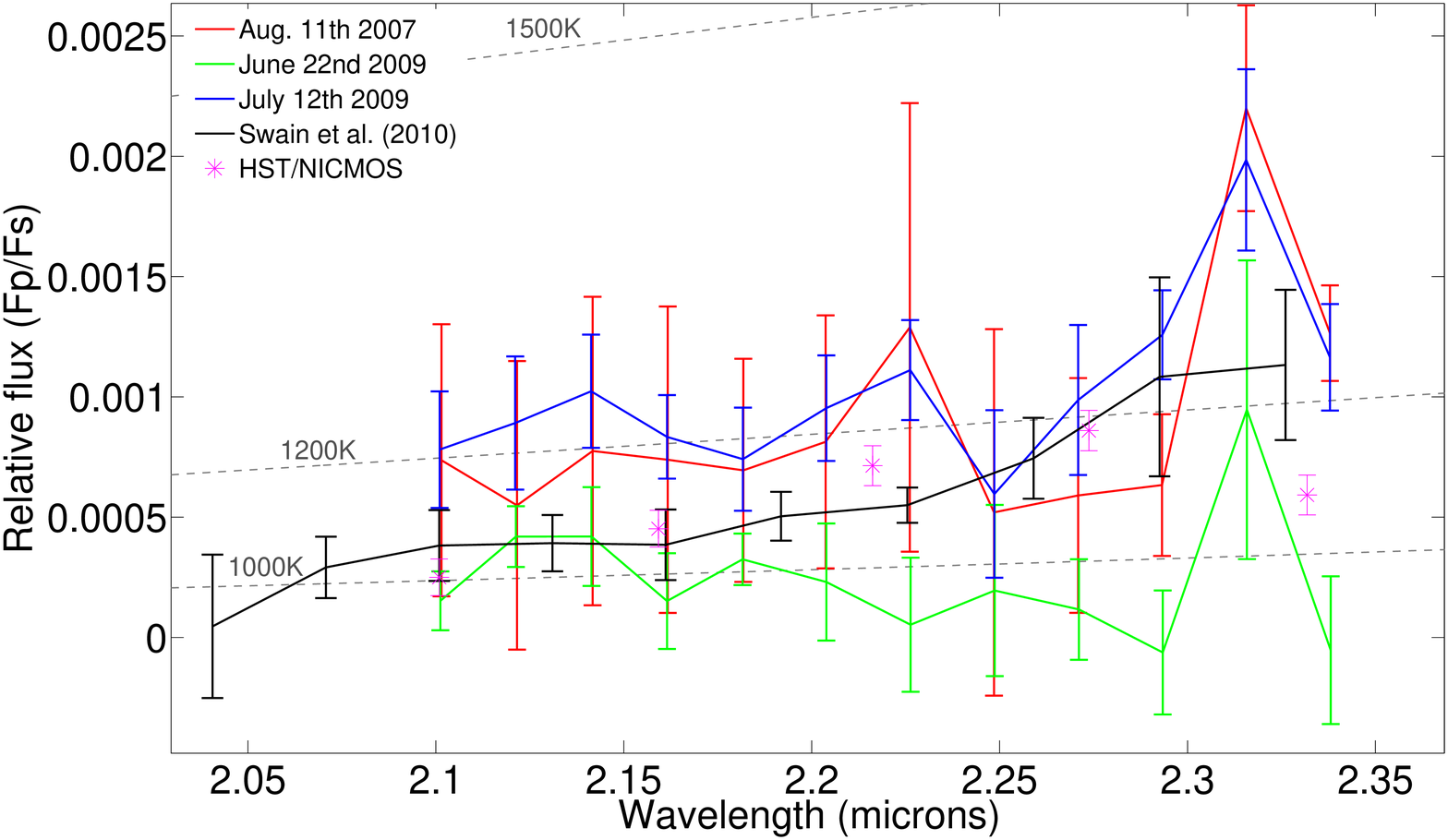}
\caption{showing K-band planetary signal for the three nights separate: August 11th 2007, June 22nd 2009 and the 12th of July 2009 in blue, red and green respectively. The night of June 22nd 2009 had poor observing conditions and the data was significantly noisier and planetary emissions retrieved are systematically lower for this night in both K and L-band. Results from \citet{swain10} are shown in black. \label{3nightsK}}
\end{figure}

Having verified the  detection of HD~189733b eclipse in all data sets, we have generated  K- and L-band spectra for each individual night as well as for all the three nights combined. 
The three individual nights are plotted in figures \ref{3nightsK} and \ref{3nightsL} for K and L bands respectively. 
All spectra are consistent with each other and are within the error bars of the initial \citet{swain10} results. 
This said, we find the nights of the 11th of August 2007 and July 12th 2009 of higher quality and in better agreement. 
The single night analysis supports the assumption that intra-night variations are negligible which allowed us to averaged the data sets and hence increase the signal to noise of the final spectra. We could hence push the resolution to R $\sim$ 170-180 for the final combined spectra. Figures \ref{gio1} and \ref{result_l} are the three-nights-combined K and L-band spectra respectively. We include in these figures the comparison with black body emission curves and LTE models. It is clear from the figures that the strong features observed in the L-band cannot be explained by standard LTE processes.
 \subsection{Comparison of the observations with atmospheric LTE and non-LTE models}
Even if many uncertainties subsist on the thermal vertical profile of HD189733b, the thermal methane emission needed to reproduce the observed spectrum would lead to brightness temperatures of $\sim$3000~K, which not only are unlikely given the star-planet configuration, but  would also appear in 
other bands --e.g. in the $\nu$4 band at 7.8 $\mu$m-- hypothesis ruled out from Spitzer observations.
While LTE models cannot explain such temperatures, non-LTE models with only stellar photons as pumping mechanism do not supply enough excess flux. This result is not unexpected since the contribution of stellar reflection from the planet is smaller in L band than the thermal emission, and fluorescence is only a redistribution of the stellar flux (even if a small enhancement comes from the redistribution of frequency in the fluorescence cascade).
However, a good fit can be obtained by assuming a vibrational temperature excess for methane by 5$\%$  due to an enhancement of the octad level population in methane which is higher than expected by stellar flux pumping (figure \ref{result_l}). This increase is currently an ad-hoc hypothesis and simply describes the amount of vibrational temperature increase required to explain the observed feature.

In the case of the K-band spectrum, it is less obvious whether LTE or non-LTE processes are prevalent. We show in figure \ref{gio1} a comparison with two LTE simulations, one including CH$_{4}$ plus CO$_{2}$ in absorption as suggested by other data sets. Another model was obtained with LTE emission of methane. However, neither of the two simulations perfectly capture the spectrum observed. 
Given the stronger non-LTE emission features detected at $\sim$3.3 $\mu$m, one can expect to find non-LTE effects in the K-band as well.  Further observations are required in order to build up the required spectral resolution to decisively constrain the excitation mechanisms at work.

\section{Discussion}

\begin{figure}
\epsscale{1.0}
\plotone{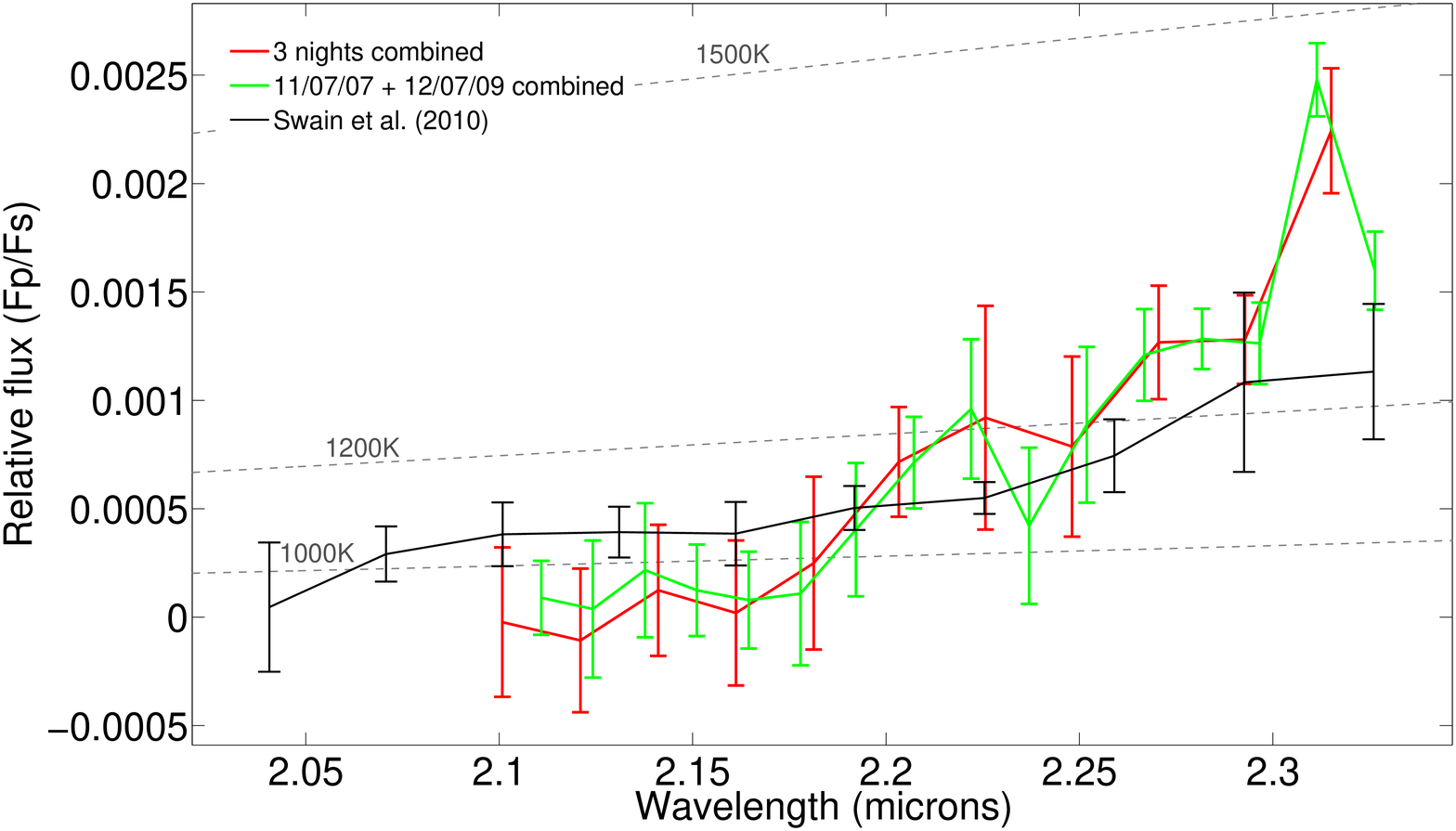}
\caption{showing the combined K-band planetary signal for the nights of August 11th 2007 and July 12th 2009 only (red), excluding the poor data quality of the June 22nd 2009 night. For comparison the spectrum of all three nights combined (green) is overplotted. The difference between both  spectra is small and indicates the night of June 22nd 2009 having a small effect on the overall result. Ground-based results from \citet{swain10} and HST/NICMOS data \citep{swain08}, are shown in black and purple respectively. \label{2nightsK}}
\end{figure}

\begin{figure}
\epsscale{1.0}
\plotone{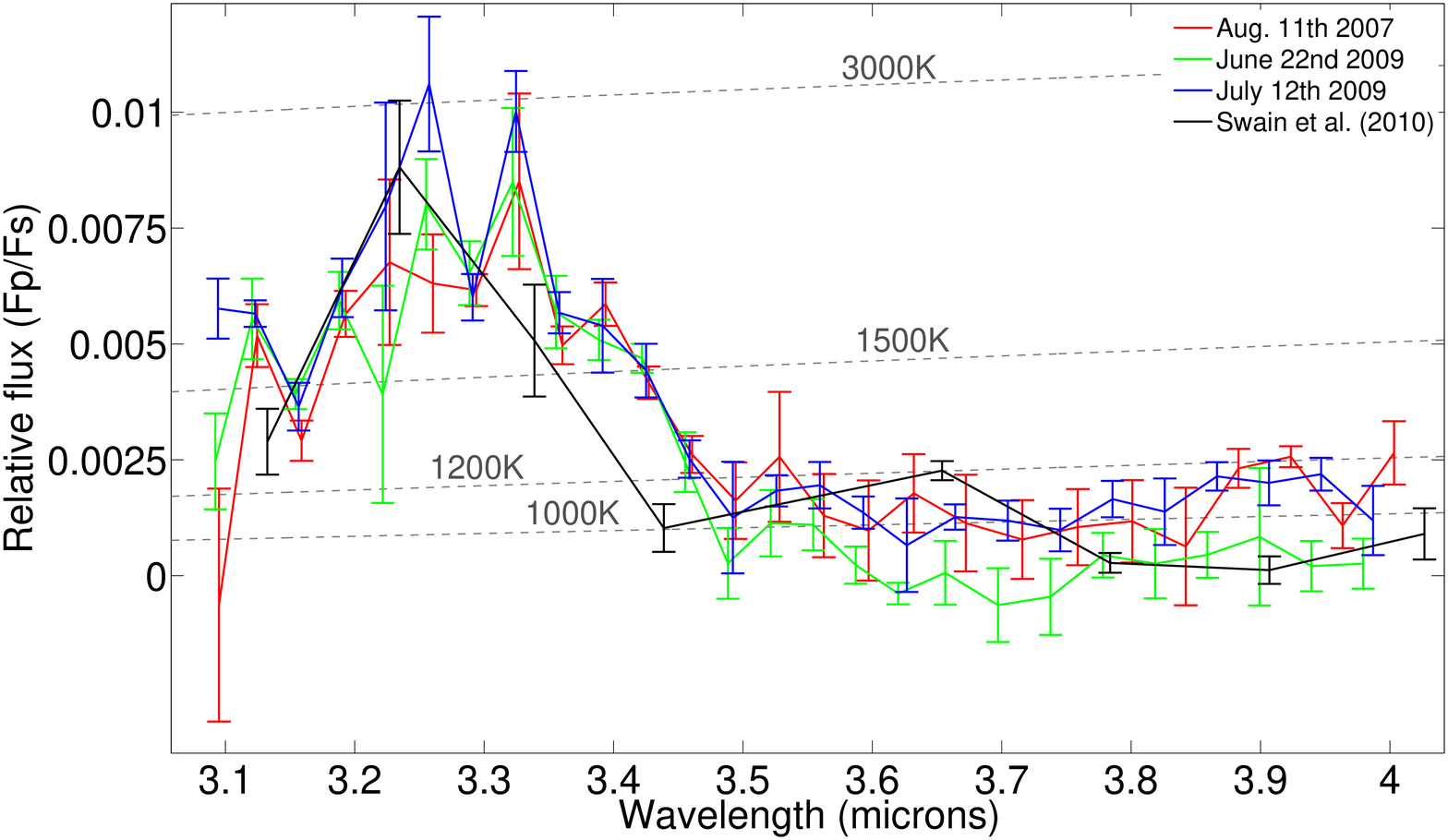}
\caption{showing L-band planetary signal for the three nights separate: August 11th 2007, June 22nd 2009 and the 12th of July 2009 in blue, red and green respectively. Similar to figure \ref{3nightsK}, the night of June 22nd 2009 shows a systematic lower emission. As described previously, this may be a result of the poor data quality of this night. Results from \citet{swain10} are shown in black. \label{3nightsL}}
\end{figure}

\begin{figure}
\epsscale{1.0}
\plotone{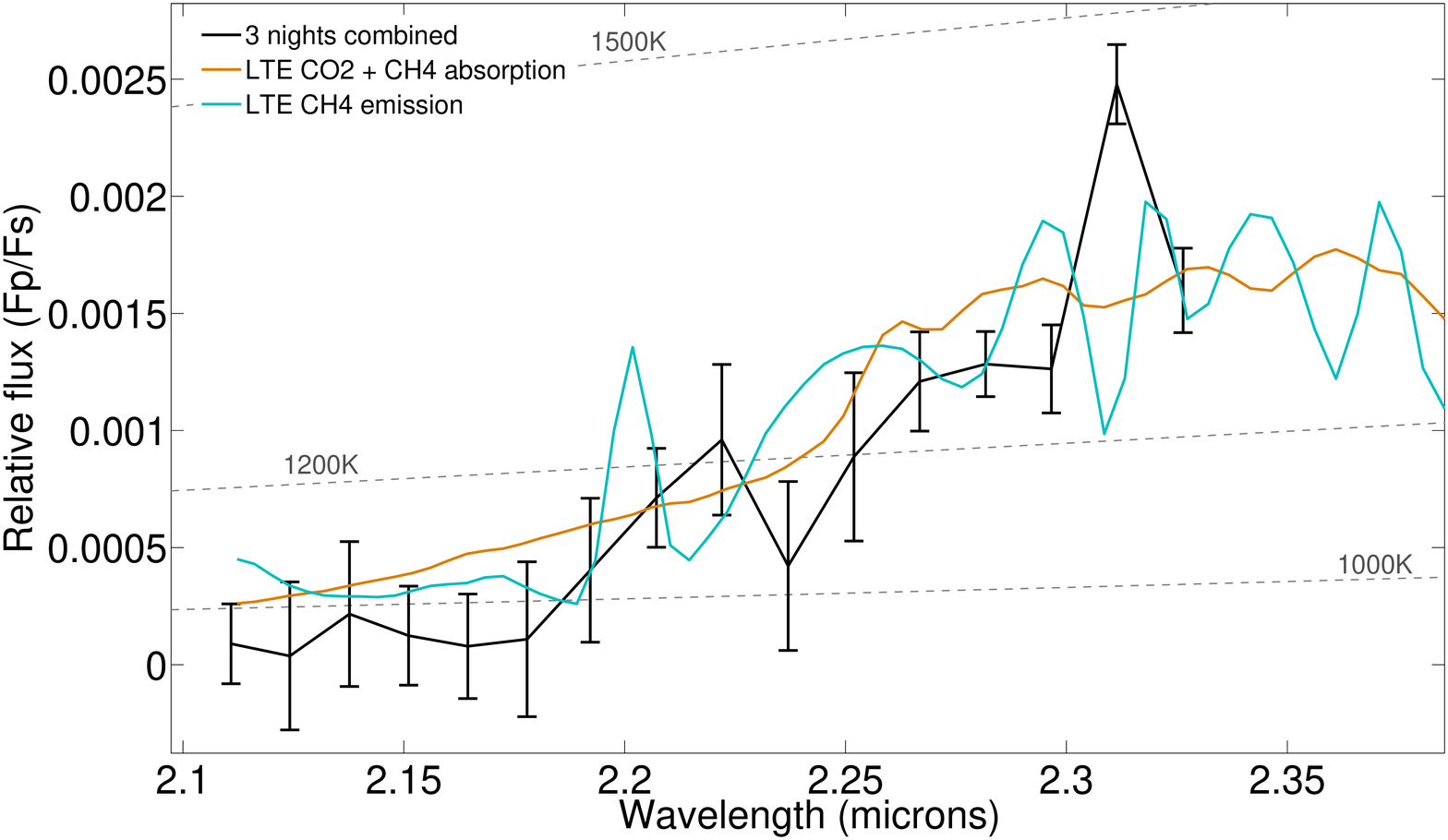}
\caption{Three night combined K band spectrum compared with three black body curves at 1000, 1500, 2000~K. Furthermore two LTE models of CH$_{4}$ in emission (turquoise) and CH$_{4}$ plus CO$_{2}$ in absorption (orange). \label{gio1} }
\end{figure}

\begin{figure}
\epsscale{1.0}
\plotone{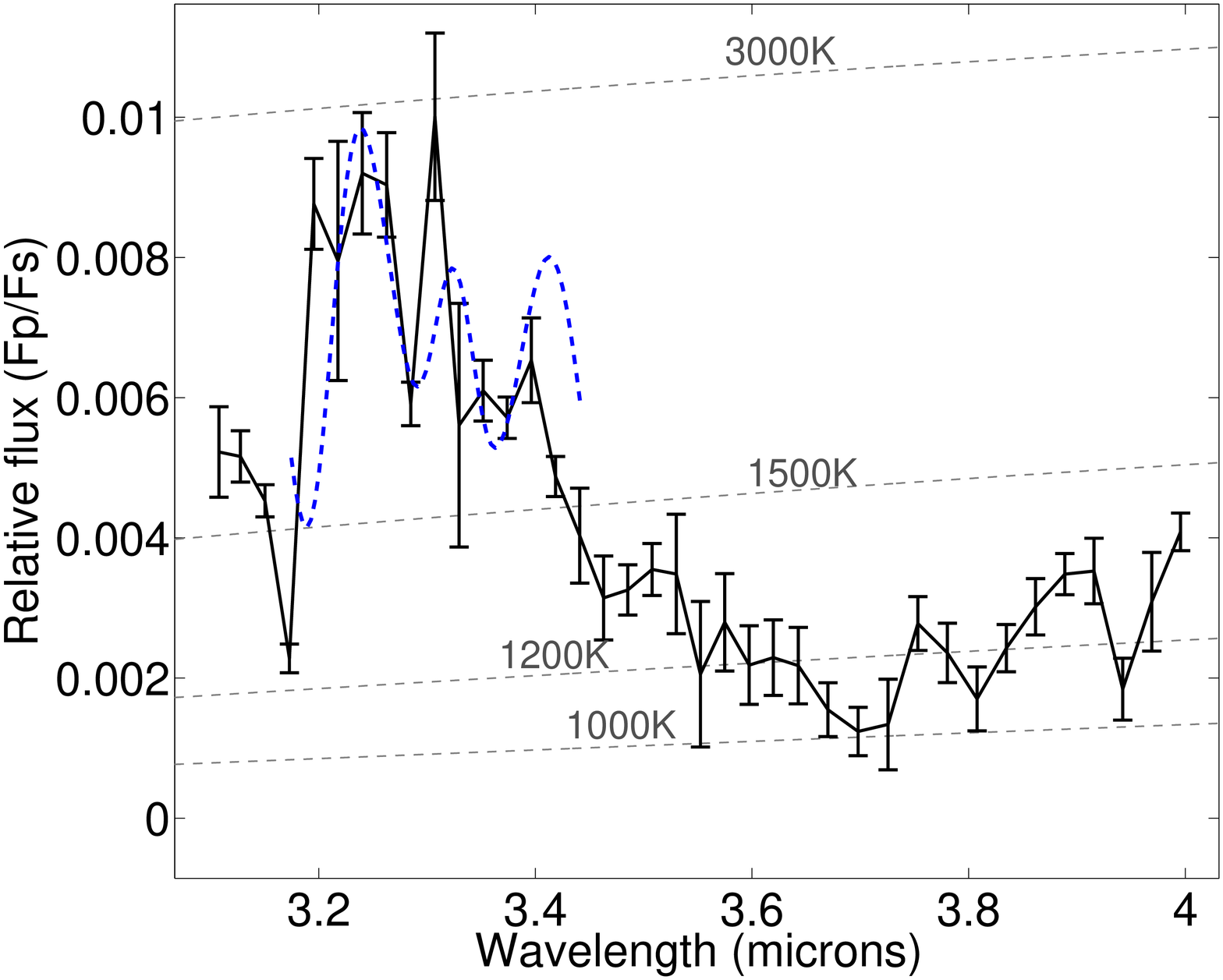}
\caption{Three nights combined L-band spectrum. The blue discontinuous line shows a comparison of the observations with the "enhanced fluorescent" model; non-thermal population enhancement in the octad level with a 5$\%$ increase of vibrational temperature of CH$_{4}$. Overlaid are black body curves at 100, 1500, 2000, 3000~K.\label{result_l}}
\end{figure}

In figure \ref{3nightsK} we present the K-band spectra of the three separate nights. This plot shows a slight discrepancy between the night of the 22nd of June 2009 compared to the other two nights analysed. We can observe a systematic off-set in both the K and L-bands (figure \ref{3nightsL}) with this night giving consistently lower emission results. We associate this effect  to the poorer observing conditions and a degraded  quality of the data compared to the data obtained in the other two nights: a very high intrinsic scatter of the data may in fact reduce the eclipse depth retrieved.  
We estimated the average spectra excluding the night of June 22nd 2009 (figure \ref{2nightsK}) and found the results to be in good agreement with the 3 nights-combined spectrum. This test demonstrates the robustness of the final retrieved spectrum. It should be noted that this issue is less severe in the L-band, since the overall signal strength is higher, than in the K-band.

Whilst the K-band spectra could be explained with  LTE  models, we encounter a quite different picture in the L-band. 
The observed emission around $\sim$3.3$\mu$m exhibits a very poor match with the predicted LTE scenario.
By contrast, non-LTE emission of methane can capture the behaviour of the $\nu_{3}$ branch. Similar fluorescence effects have been observed in our own solar system, mainly CO$_2$ in telluric and CH$_4$ in giant solar system planets \citep{barthelemy05}. In section \ref{non-LTE} we outline a plausible model for the creation of such a prominent feature. As previously mentioned, the increase in CH$_{4}$ vibrational temperature of 5$\%$ is presently an ad-hoc hypothesis: it simply describes the amount of non-LTE population required to fit the observations, pure LTE populations being insufficient.  The source of this population increase can come for a variety of sources: XUV illumination from the star, electron precipitations, etc.which are presently not constrained at all. Such effects are nonetheless known in planetary physics, such as on Jupiter, where H$_{2}$ vibrational temperatures in the upper atmosphere have been demonstrated to be out of equilibrium through Ly-alpha observations \citep{barthelemy05}, with a 1.4-1.5 fold increase in vibrational temperature.


\subsection{Validation of observations}
\label{contamination}

The results presented here are found to be consistent with the results initially presented by \citet{swain10}, HST/NICMOS data in the K-band \citep{swain08} and verified in the L-band by the Spitzer/IRAC 3.6$\mu$m broadband photometry \citep{charbonneau08}, see figures \ref{2nightsK} $\&$ \ref{3nightsL}. However, \citet{mandell11}, from here M11, recently published a critique of the original \citet{swain10}, from here S10, result reporting a non-detection of any exoplanetary features in their analysis. Since the results of this publication are in good accord with \citet{swain10}, the fundamental discrepancy between the findings presented here and those by M11 need to be addressed. 

M11 argue that the L-band features reported by S11 were likely due to un-accounted for telluric water emissions rather than exoplanetary methane. This hypothesis poses four main questions which will be addressed below: (1) Do the L-band features look like water emissions? (2) Are the results repeatable? (3) Do or do we not see similar lightcurve features in the reference star? (4) Can we quantify the amount of residual telluric contamination in the data? 



\subsubsection{Do the L-band features look like water?}

Here the simple answer is no. As discussed in section \ref{results} and shown in figures \ref{3nightsL} $\&$ \ref{result_l}, the improved spectral resolution of these results shows that we are clearly dealing with methane signatures. 
As M11 pointed out, a temporary change in telluric opacity due to atmospheric water (or methane) could mimic a secondary eclipse event. However, for temporal atmospheric variations to mimic an eclipse signal in the combined result of all three nights, the opacity variations, as well as the airmass function, would need to be identical or at least very similar in all data sets. The likelihood of such hypothesis  is very small.  

In addition, we have retrieved weather recordings from near-by weather stations. These include periodic temperature, relative humidity and pressure readings from the the CFHT\footnote{http://mkwc.ifa.hawaii.edu/} as well as atmospheric opacity (tau) readings at 225 $\mu$m obtained by the CSO\footnote{http://www.cso.caltech.edu/} (see figure \ref{weather}). Spread over all three eclipsing events, we found no significant correlations between these parameters and the secondary transit shape expected.

\subsubsection{Are the results repeatable?}

A main focus throughout this publication is to demonstrate the repeatability of the observations. In section \ref{results} we present spectra retrieved for each individual observing run of the three 'short-slit'  nights and found them consistent with each other within the error-bars. For the methane $\nu_{3}$ band which is the most difficult to achieve measurement we present lightcurves for all three observing runs considered, figure \ref{3nightnu3}. These do vary in quality from night to night but are found to be consistent with one another over a measured timescale ranging from August 11th 2007 to July 12th 2009. This test of repeatability is of paramount importance in asserting the validity of the analysis as a whole.

\subsubsection{ Do or do we not see similar lightcurve features in the reference star? }

\begin{figure}
\epsscale{1.0}
\plotone{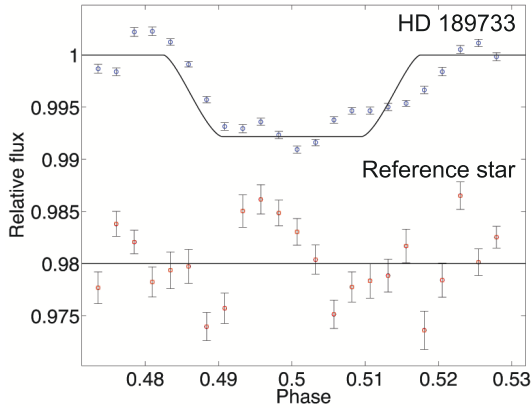}
\caption{showing the lightcurves of the long-slit analysis of HD189733b and the simultaneously observed fainter reference star beneath, centred at 3.31$\mu$m with the standard 50 channel binning. Overplotted are two fitted \citet{mandel02} curves for the secondary eclipse. The HD189733b lightcurve is in good agreement with the other results of this paper whilst the reference star's timeseries is noticeably flat. \label{mstar}}
\end{figure}

\begin{figure}
\epsscale{1.0}
\plotone{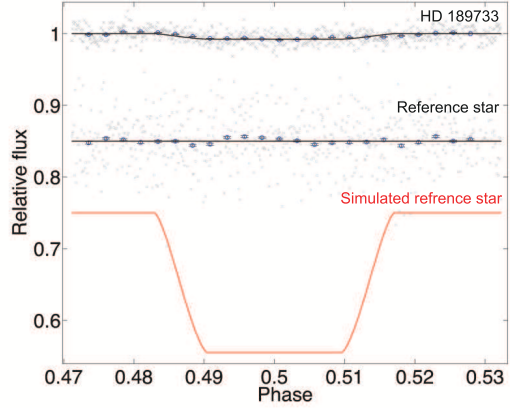}
\caption{showing on the top the observed lightcurve of HD189733b, beneath the simultaneously observed flat timeseries of the fainter reference star. At the bottom in red is the simulated reference star lightcurve expected to be observed under the assumption that the observed signal in HD189733b is due to an imperfect background subtraction. The flat nature of the observed reference star lightcurve is a strong indication that the background subtraction was treated adequately.  \label{mstar2}}
\end{figure}

\begin{figure}
\epsscale{1.0}
\plotone{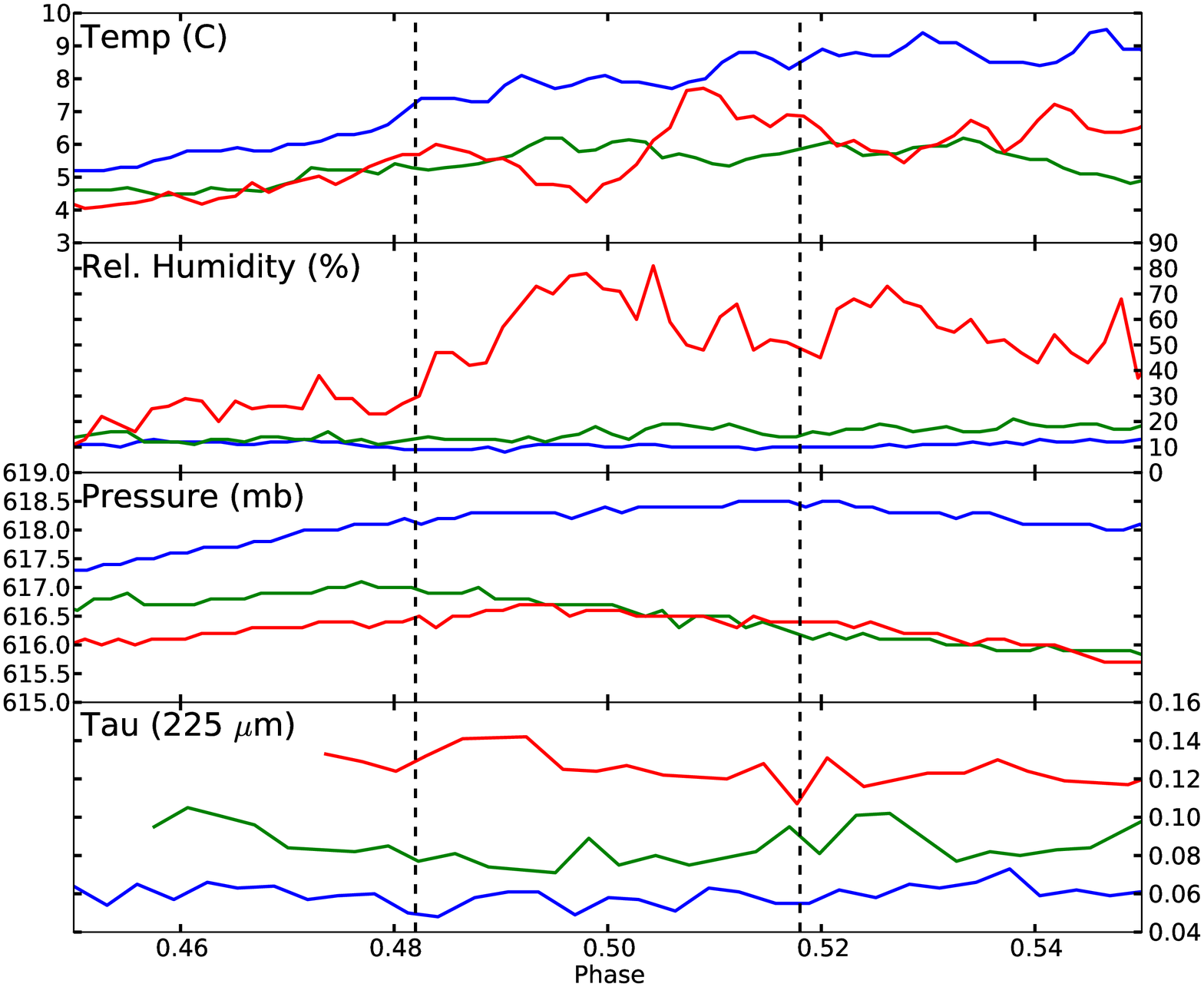}
\caption{showing from top to bottom: Temperature (deg. C, CFHT Weather station), Rel. Humidity ($\%$,CFHT), Pressure (mb,CFHT) and optical depth, tau (225$\mu$m, CSO) for the 12nd Aug. 2007 (blue), 22nd June (green) and 12th July 2009 (red). The discontinuous vertical lines mark the secondary transit duration.}
\label{weather}
\end{figure}

We do not see any lightcurve features in the reference star's timeseries. As described in previous sections, we have obtained a fourth night in addition to the three main nights analysed here. This fourth night was taken in the single-order, L-band only mode with a one arc-minutes long slit. This allowed us to simultaneously observe the target HD189733b and a fainter reference star, 2MASS 20003818+2242065, over the course of a secondary eclipse on July 3rd 2010. We have equally applied the same routines outlined in section \ref{specextract} to both, the target and the reference.  In figure \ref{mstar} we plot the resulting lightcurves of both stars centred at 3.31$\mu$m using the standard 50 channel bin. We find the transit depth for HD189733b to be within the error bars of the other nights analysed, whilst the reference star timeseries is flat. Hence, the routines used produce a null result where a null detection is expected. 

Furthermore it is important to note that a faulty background subtraction would have much stronger effects on the fainter reference star than on the target, as any residual background is a proportionally larger fraction of the stellar signal. We find the mean observed flux for a single exposure to be  F$_{HD189} \sim$24300e$^{-}$  and F$_{REF} \sim$900e$^{-}$ for the target and the reference stars respectively. We can now state that the observed flux is a sum of the stellar flux and a background contribution: F$_{observed}$ = F$_{star}$ + F$_{back}$. We also assume that the background flux, F$_{back}$, is the same for both stars as they were observed simultaneously on the same detector. Whatever the value of F$_{back}$ may be, its relative contribution on the overall flux would be $\sim$27 times higher for F$_{REF}$ than for F$_{HD189}$. Following this argument, if we now assume the lightcurve feature to be due to an inadequate background correction (as postulated by M11), we would expect a $\sim$27 times deeper lightcurve signal in the reference star timeseries than in HD189733b. To illustrate the severity of this effect, we re-plotted the timeseries presented in figure \ref{mstar} with an additional 27 times deeper transit than that of HD189733b underneath. Given the flat nature of the reference star's timeseries though, we can confidently confirm an adequate treatment of telluric and other backgrounds.

\subsubsection{Can we quantify the  residual telluric contamination in the data? }

\begin{figure}
\epsscale{1.0}
\plotone{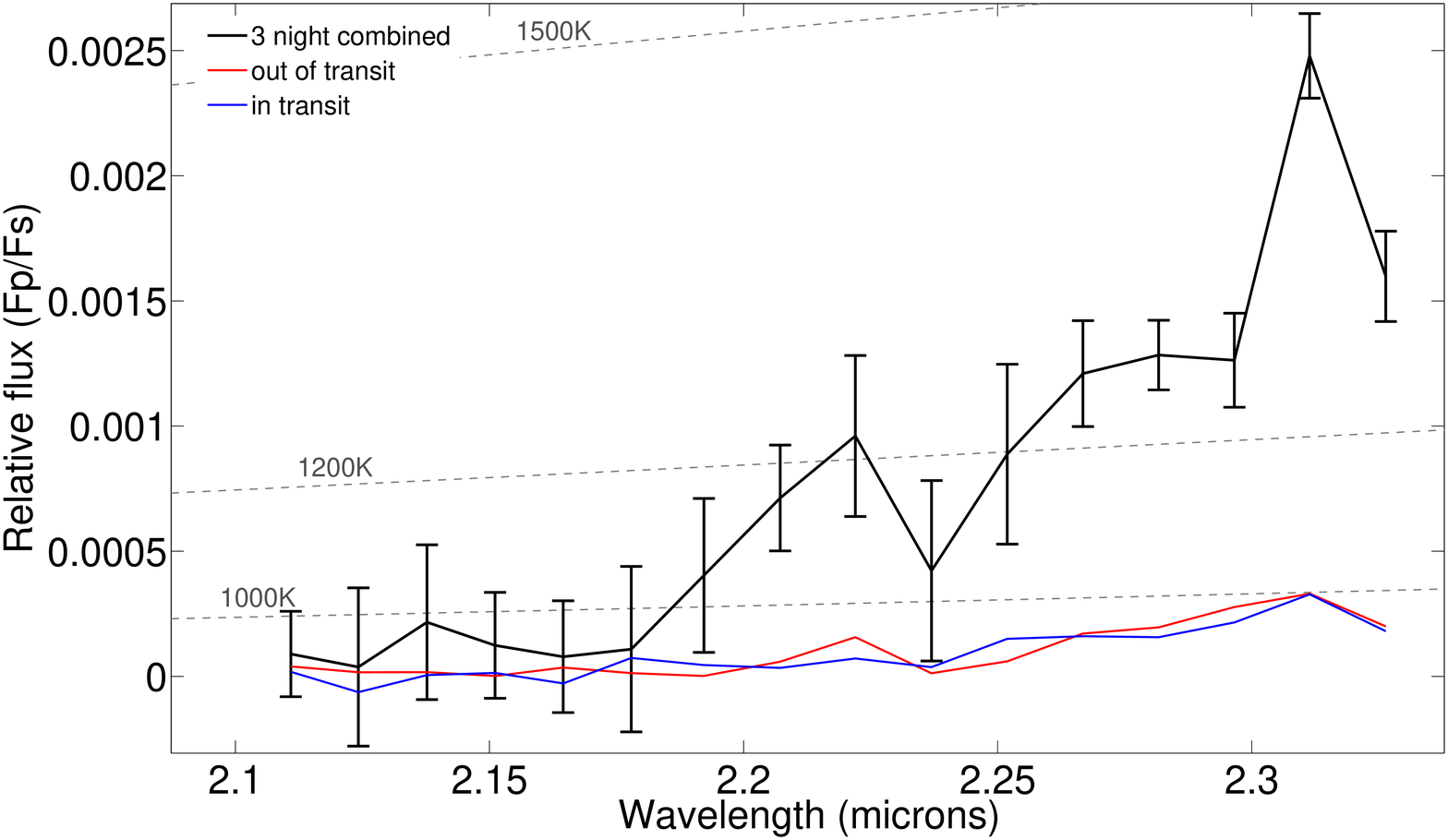}
\caption{showing the three night combined K-band result (black), in-transit and out-of-transit contamination measures are plotted in blue (dash-dotted) and red (dashed) respectively. It can clearly be seen that the contamination by telluric components is much smaller than the planetary signal and the it's amplitude lies within the signal's error bar.\label{contamK}}
\end{figure}
\begin{figure}
\epsscale{1.0}
\plotone{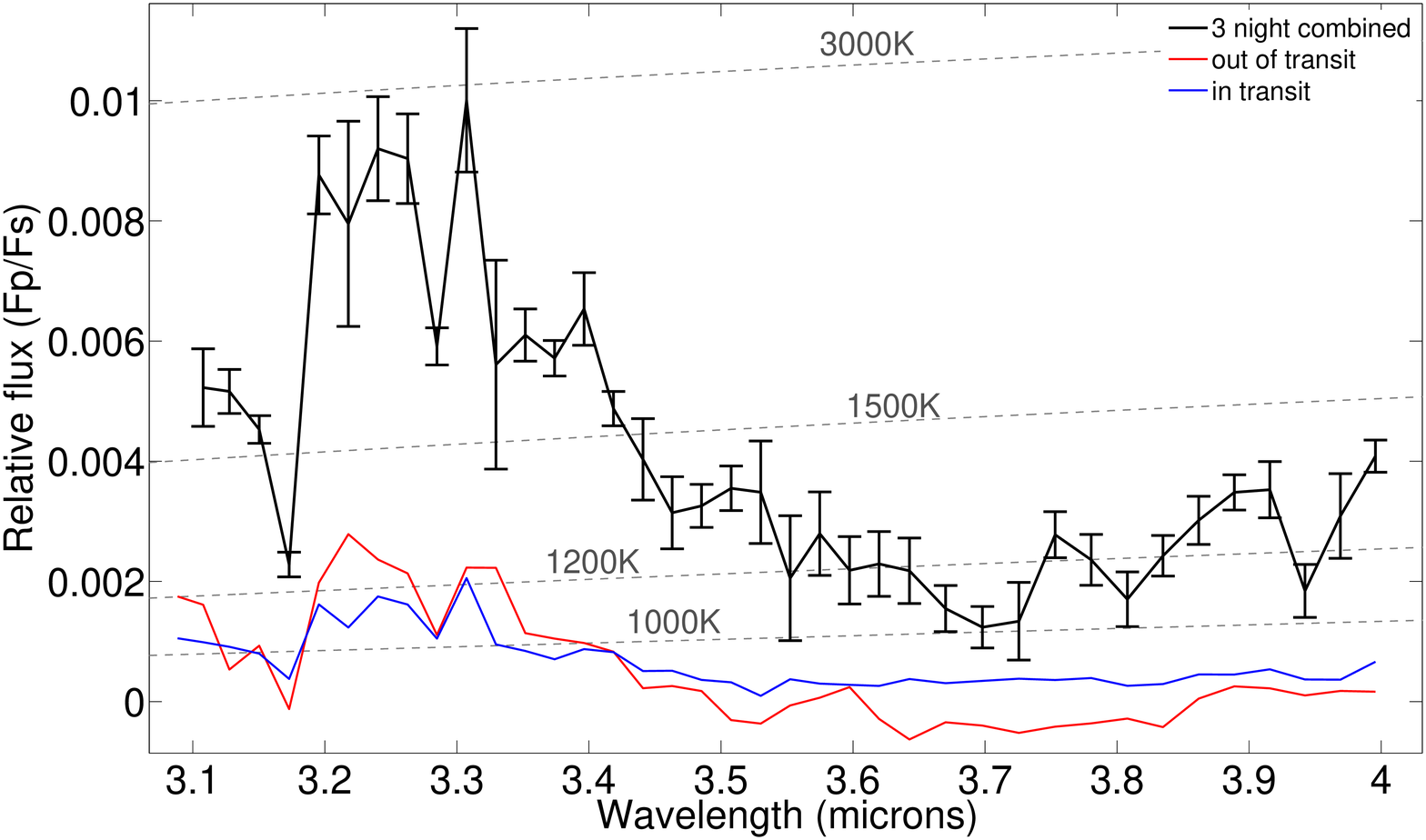}
\caption{showing the three night combined L-band result (black), in-transit and out-of-transit contamination measures are plotted in blue (dash-dotted) and red (dashed) respectively. It can clearly be seen that the contamination by telluric components is much smaller than the planetary signal and the it's amplitude lies within the signal's error bar.\label{contamL}}
\end{figure}

Using the Fourier based techniques described in this paper, we can quantify the remaining contribution of systematic noise and the residual telluric components in the spectra shown in sec. \ref{results}. As described in section \ref{fourier}, we are mapping individual Fourier coefficients of the lightcurve signal in the frequency domain. Any systematic noise or telluric contamination can therefore only contribute to this one frequency bin. The degree of residual contamination by systematics on that frequency bin can hence be estimated by running the  routine  described in section \ref{frequdomain} on only out-of-transit and only in-transit data, i.e. removing the eclipse signal. Figure \ref{contamK} and \ref{contamL} show the planet signal (black) and out-of-transit and in-transit measurements of the contamination in red and green respectively. We conclude that the amplitude of the systematic noise and the residual telluric component is within the error bars of the planetary signal. 

\section{Conclusion}

In this paper we present new data on the secondary eclipse of HD~189733b recorded with the SpeX instrument on the IRTF.  
Our data analysis algorithm   for time-resolved, ground-based spectroscopic data, is based on a thorough pre-cleaning of the raw data and subsequent spectral analysis using Fourier based techniques.  
By combining three nights of observations, with identical settings, and a further development of the data analysis methodology presented in \citet{swain10}, we could to increase the spectral resolution to R $\sim$ 175. 

We confirm the existence of a strong feature at $\sim$ 3.3$\mu$m, corresponding to the methane $\nu_{3}$ branch, which cannot be explained by  LTE models. Non-LTE processes are most likely the origin of such emission and we propose a plausible scheme to explain it. 

The possibility of telluric contamination of the data is thoroughly tested but 
we demonstrate that the residual due to atmospheric leakage is well within the error-bars, both by using Fourier based techniques and additional observations with a reference star in the slit. 
This critical test demonstrates the robustness of our calibration method and its broad applicability in the future to other space and ground exoplanet data.

\acknowledgments
I.P.W. is supported by a STFC Studentship. We like to thank the IRTF, which is operated by the University of Hawaii under Cooperative Agreement no. NNX-08AE38A with the National Aeronautics and Space Administration, Science Mission Directorate, Planetary Astronomy Program.

%

\appendix
\section{Additional notes on wavelets}

As mentioned in section \ref{secwave}, wavelet de-noising of timeseries data has several advantages: 
 1) wavelet de-composition is a non-parametric algorithm and hence does not assume prior information on the signal or noise properties, making it an easy to use and objective de-noising routine; 2) contrary to smoothing algorithms (eg. kernel regression) high and low signal frequencies are retained; 3) temporal phase information of the signal is preserved during the de- and re-construction of the signal. This allows for an optimal white and systematic noise reduction at varying frequency pass-bands.  For our purposes we use a non-linear wavelet shrinkage by soft-thresholding of the obtained wavelet coefficients and iterative reconstruction of the data. The intricacies of such an approach were extensively discussed by \citet{donoho95} and \citet{persival00}. Using the 'Wavelet Toolbox' in MATLAB, each individual timeseries underwent a 4 level wavelet shrinkage using "Daubechies 4" wavelets. The wavelet coefficients were estimated for each decomposition step using an heuristic form of the Stein's Unbiased Risk Estimate (SURE) for soft-thresholding \citep{stein81}. This allows for a MINMAX coefficient estimation \citep{sardy00} in cases of too low signal-to-noise (SNR) for the SURE algorithm.  After thresholding, the timeseries were reconstructed based on the obtained coefficients for each timeseries. 

\section{Fourier analysis}

In section \ref{frequdomain}, we discuss the properties of box-shaped lightcurves in the frequency domain. It is needless to say that this is a gross over-simplification and that the actual secondary eclipse lightcurve is more akin to a trapezoid (equations \ref{trapez}) rather than a square-box. In the case of a trapezoid, we can calculate the power to decrease by $1/k^2$ for Fourier coefficients above k=1. Hence, the Fourier series for a trapezoidal shape converges faster (equation \ref{trapezpow}). 

\begin{align}
\label{trapez}
f_{trap}(t)  &=  8\sqrt(2)\delta \left( \text{sin}(1/\tau) + \frac{\text{sin}(3/\tau)}{9}  - \frac{\text{sin}(5/\tau)}{25} -  ... \right)  \\\nonumber
 &=  8\sqrt(2)\delta \sum_{k=1,3,5...}^{\infty} \left(\frac{\text{sin}(k/4\tau) + \text{sin}(3k/4\tau)}{k^2} \right)
\end{align}

\begin{equation}
\label{trapezpow}
|A|_{trapez}= \frac{\tau \delta}{2} \sum_{k=1,3,5...}^{\infty} \frac{1}{k^2} 
\end{equation}

The difference between the box-car and trapezoidal shape do not affect the linear relationship between spectral amplitude and transit depth. We furthermore extend the argument to limb-darkened lightcurves that exhibit a markedly rounder morphology. These are a natural extension to the trapezoidal case and it is generally true that the 'rounder' the eclipse shape, the less power is contained in Fourier coefficients above $k$ = 1, and hence the series are converging even faster.

\bibliographystyle{apj}
\bibliography{exobib}

\end{document}